\documentclass[preprint, 3p, twocolumn]{elsarticle}
\usepackage{graphicx}
\usepackage{booktabs}
\usepackage{tabularx}
\usepackage[normalem]{ulem}
 \useunder{\uline}{\ul}{}
 \usepackage{rotating}
 \usepackage{dsfont}
 \usepackage{textcomp}
\usepackage{gensymb}
\usepackage{color}
\def\Manu#1{\textcolor{black}{#1}}

\usepackage{graphicx,xcolor}
\usepackage{epstopdf}
\usepackage{setspace}
\usepackage[T1]{fontenc}
\usepackage{amssymb,amsfonts,amsmath}
\usepackage{setspace}
\usepackage{subcaption}
\usepackage{mathtools}
\usepackage{etoolbox}
\newcolumntype{Y}{>{\centering\arraybackslash}X}
\usepackage[switch, mathlines]{lineno}
\usepackage{upgreek}

\let\oldequation\equation
\let\oldendequation\endequation

\renewenvironment{equation}
  {\linenomathNonumbers\oldequation}
  {\oldendequation\endlinenomath}

\makeatletter
\def\ps@pprintTitle{%
 \let\@oddhead\@empty
 \let\@evenhead\@empty
 \def\@oddfoot{\footnotesize\itshape
       Preprint submitted to \ifx\@journal\@empty Physics of Fluids. 
       \else\@journal\fi\hfill\today}
 \let\@evenfoot\@oddfoot}
\makeatother

\usepackage{hyperref}
\hypersetup{
    colorlinks=true,
    linkcolor=blue,
    filecolor=magenta,      
    urlcolor=cyan,
}

\begin{document}

\title{\Manu{Influence of Salt on the Formation and Separation of Droplet Interface Bilayers}} 
\author[yogi]{Y. Huang}
\author[yogi,vinny,wyss]{V. Chandran Suja}
\author[yogi]{L. Amirthalingam}
\author[yogi]{G.G. Fuller}
\ead{ggf@stanford.edu}
\address[yogi]{Department of Chemical Engineering, Stanford University, Stanford, California 94305}
\address[vinny]{School of Engineering and Applied Sciences, Harvard University, MA - 01234, USA}
\address[wyss]{Wyss Institute for Biologically Inspired Engineering, 52 Oxford St, Cambridge, MA 02138, USA}
\begin{abstract}
Phospholipid bilayers are a major component of the cell membrane that is in contact with physiological electrolyte solutions including salt ions. The effect of salt on the phospholipid bilayer mechanics is an active research area due to its implications for cellular function and viability. In this manuscript we utilize droplet interface bilayers(DIBs), a bilayer formed artificially between two aqueous droplets, to unravel the bilayer formation and separation mechanics with a combination of experiments and numerical modelling under the effects of K$^+$, Na$^+$, Li$^+$, Ca$^{2+}$ and Mg$^{2+}$. Initially, we measured the interfacial tension and the interfacial complex viscosity of lipid monolayers at a flat oil-aqueous interface and show that both properties are sensitive to salt concentration, ion size and valency. Subsequently, we measured DIB formation rates and show that the characteristic bilayer formation velocity scales with the ratio of the interfacial tension to the interfacial viscosity. Next, we subjected the system to a step strain by separating the drops in a stepwise manner. By tracking the evolution of the bilayer contact angle and radius, we show that salt influences the bilayer separation mechanics including the decay of the contact angle, the decay of the bilayer radius and the corresponding relaxation time. Finally, we explain the salt effect on the observed bilayer separation by means of a mathematical model comprising of the Young-Laplace equation and an evolution equation.
\end{abstract}


\maketitle
\section{Introduction}
The phospholipid bilayer is a crucial component of the cell membrane that serves as a barrier for molecular and ion regulation \cite{lombard2014once,naumann2008protein,bello2017lipid}. It was first identified by Gorter and Grendel \cite{gorter1925bimolecular}, and then the discovery of the fluid mosaic model of the cell membrane by Singer and Nicolson paved the way for development of the physicochemical characteristics of the lipid bilayer \cite{singer1972fluid}. Physicochemical characteristics of the phospholipid bilayers are influenced by salt ions in nearby electrolyte solutions, which is physiologically in contact with the cell membrane \cite{lin2016effects}. Common salt ions, such as K$^+$, Na$^+$, Li$^+$, Ca$^{2+}$ and Mg$^{2+}$, can play pivotal role to the cell membranes in human body to regulate human brain or control fatal diseases \cite{somjen2002ion,ratner2012immune}. One aspect of the salt influence on the bilayer relates to the mechanics of ion binding to the lipid bilayer \cite{maity2016binding,redondo2014structural}. Another important aspect relates to the influence of salt on the force dynamics within the bilayer, which is explored in the present work, first through the measurement of interfacial tension and interfacial complex viscosity of lipid monolayers, and then through the formation and separation of the bilayer. 

\begin{figure*}[!h]
\centering
\includegraphics[width=\linewidth]{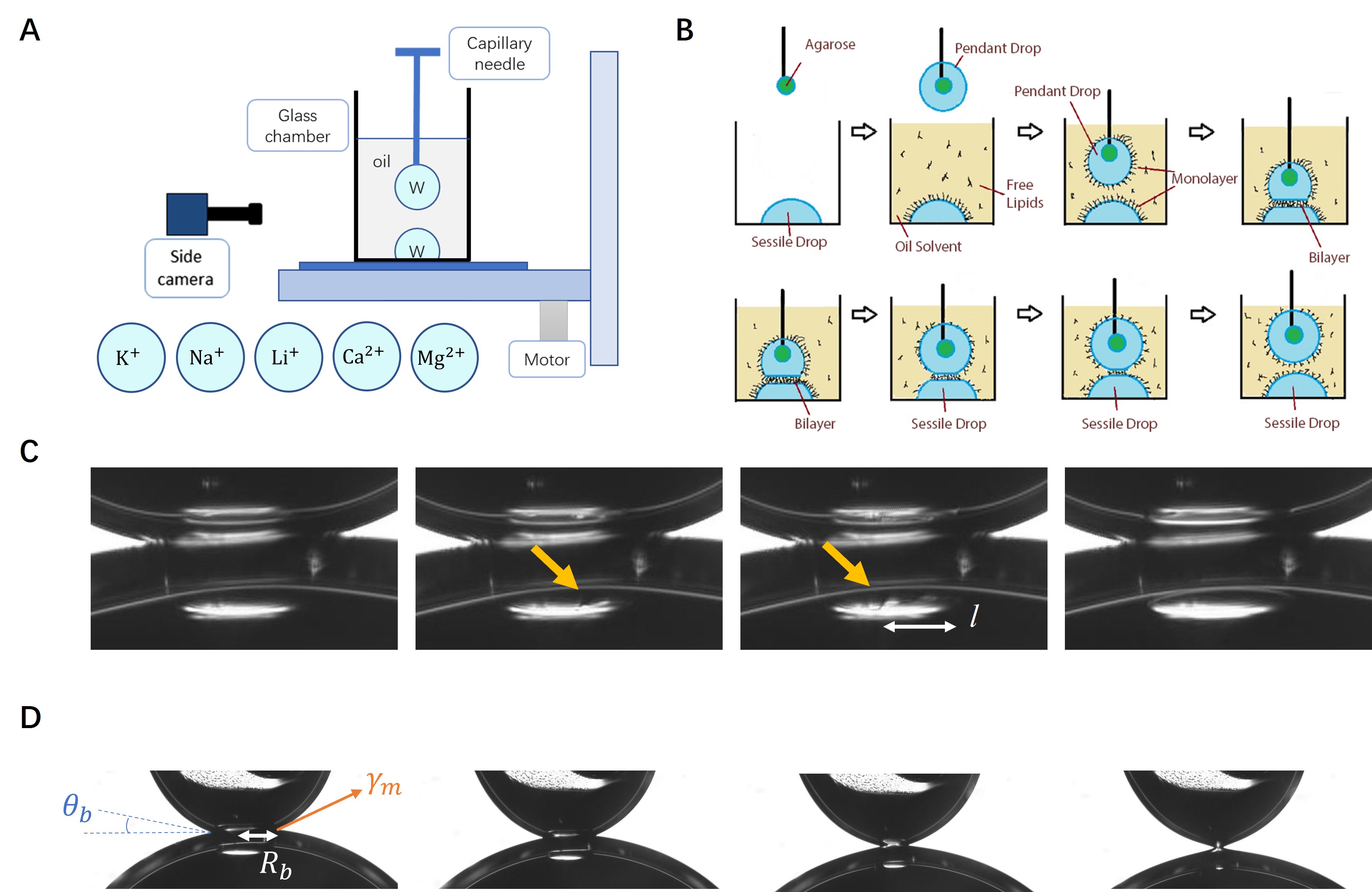} \caption{A schematic of the experimental setup and protocol. (A)Schematic of the Interfacial Drainage Dilatational and Stability Stage (I-DDiaSS) setup along with the labeled components. Five salt cations (K$^+$, Na$^+$, Li$^+$, Ca$^{2+}$ and Mg$^{2+}$) included respectively as a component of the pendant and sessile droplet. (B)Experimental protocol for bilayer formation and separation: a sessile droplet of predetermined salt solution is pipetted on the bottom of the chamber. A pendant droplet of the same salt solution is pipetted to a capillary with an agarose core at the tip. 0.2 mL of a 10 mM solution of DPhPC lipid in hexadecane is added into the chamber and then the pendant droplet is submerged. Aged for 10 minutes to allow the formation of the lipid monolayer on the droplet interface, the droplets are brought into contact. Then the bilayer is formed after few minutes ($\sim5\mbox{\,min}$) and then the droplets are separated in a step wise manner until the bilayer is finally separated. (C)A sequence of images obtained from the side camera using 1M KCl salt solution showing the formation of the bilayer. The yellow arrows indicate the advancing front of the bilayer as it forms between the two droplets. The position of the front line is denoted as $l$. (D)Sequence of images using 1M KCl salt solution showing the bilayer at the beginning of each step and just before separation. $\theta_b$ is the contact angle, $R_b$ is the bilayer radius and $\gamma_m$ is the monolayer surface tension at the droplet contact line. Images in (A) and (B) are adapted with permission from ref \cite{huang2021surface}.} \label{fig:Setup}
\end{figure*}

The techniques of fabricating $in$ $vitro$ phospholipid bilayers include solid supported lipid bilayers (SLBs), giant uni-lamellar vesicles (GUVs), black lipid membranes (BLMs) and droplet interface bilayers (DIBs) \cite{beltramo2016millimeter,funakoshi2006lipid,evans1987physical}. Droplet interface bilayers, the bilayers formed between two aqueous droplets within non polar lipid solutions, are an attractive format that has been commonly used to study the physical \cite{dixit2012droplet,venkatesan2015adsorption,yanagisawa2013adhesive}, electrical \cite{hwang2007electrical,punnamaraju2011voltage} and transport characteristics \cite{lee2018static,fleury2020enhanced} of cell membranes. The key advantage of droplet interface bilayers is that the drop profiles around the bilayers can be conveniently visualized by cameras, which allows one to track the mechanical response of the drops and the bilayers subjected to external forcing \cite{taylor2015direct,najem2015activation,najem2017mechanics,huang2022physicochemical}. Recently, our group reported the separation mechanics of DIBs under step strains \cite{huang2021surface}. Combining the experiments and numerical modelling, we have shown that the droplets separate primarily through a peeling process with the dominant resistance to separation coming from viscous dissipation associated with corner flows.


\begin{figure*}[!h]
\centering
\includegraphics[width=\linewidth]{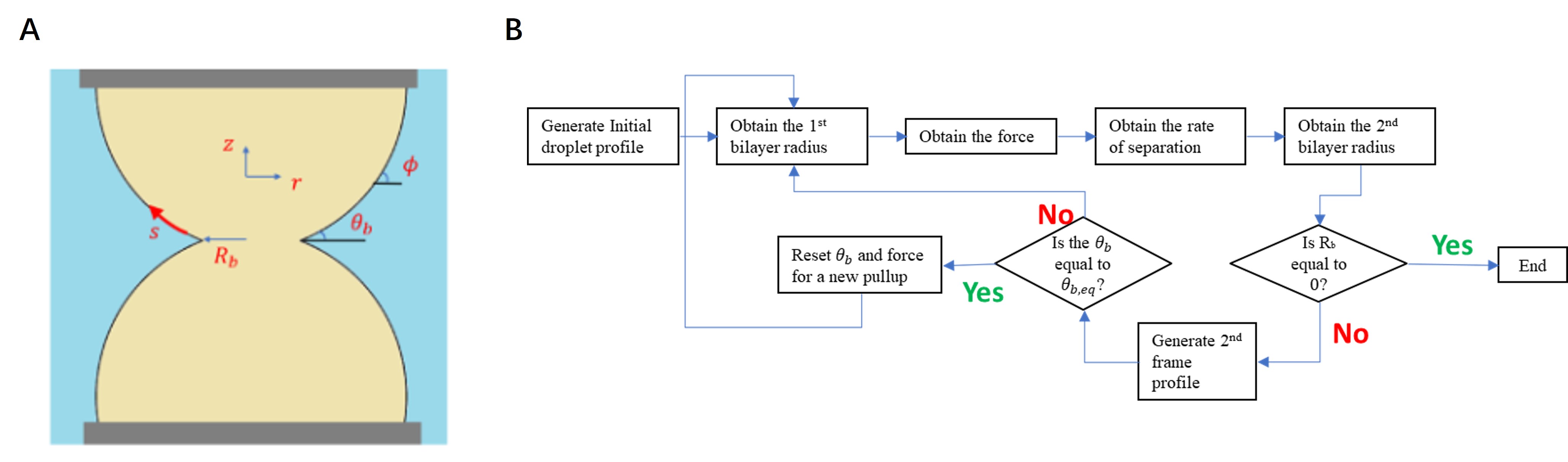} \caption{(A) A schematic of the pendant-sessile drop system used in the mathematical model. (B) Flow chart showing the iterative solution of the equations governing the bilayer separation. Images in (A) and (B) are adapted with permission from ref \cite{huang2021surface}.} \label{fig:SimFlowChart}
\end{figure*}

Despite the vast literature on the binding mechanics of the salts onto the lipid molecules, salt effects on bilayer separation mechanics (unzipping) has not been previously considered. We address this knowledge gap in this manuscript and study the effect of K$^+$, Na$^+$, Li$^+$, Ca$^{2+}$ and Mg$^{2+}$ ions on DIBs through a combination of experiments and mathematical modeling. First, we measure the interfacial tension (IFT) and complex viscosities of lipid monolayers at flat oil-aqueous interfaces. Second, we create DIBs subject to strain mechanics which form the basement for the mathematical model for the detachment processes. The key findings from this study are reported in Section \ref{sec:results}, where we show (i) the IFT and complex viscosities of lipid monolayers with the tested salts in the aqueous phase, (ii) experimental data of the bilayer formation and separation dynamics that highlight how IFT and complex viscosity can influence the formation and separation mechanics following step strain, and (iii) a quantitative comparison of the experimental data against the predictions of bilayer separation using this mathematical model. 

\section{Materials and Methods}
\subsection{Materials}
Salt, namely KCl, NaCl, LiCl, CaCl$_2$ and MgCl$_2$, were purchased from Sigma-Aldrich and dissolved with distilled water to give sample salt solutions. 1,2-diphytanoyl-sn-glycero-3-phosphocholine (DPhPC) solution in chloroform was purchased from Avanti Polar Lipids (Catlog no: 850356) and was used to generate lipid monolayers and bilayers. Agarose powder was purchased from Thermo Fisher Scientific (Catlog no: BP164100) and was used to support the pendant drop (Fig. \ref{fig:Setup}B).

Prior to the experiments DPhPC was extracted by blowing a nitrogen stream for 45 minutes to evaporate off the chloroform solution. Subsequently the residual lipid film was vacuum dried for 60 minutes and then dissolved in hexadecance (hereafter referred as oil) to give a concentration of 10 mM. 300 mg of the agarose powder was mixed with 10 mL of distilled water at high temperature, and then cooled down to make the agarose gel~\cite{holden2007functional,leptihn2013constructing}. A predetermined amount of salt solutions were used for preparing the aqueous sessile and pendant droplets. The agarose core size at the tip of the capillary was ensured to be much smaller than the pendant drop size to avoid any undesired influence of the agarose core on the DIB dynamics.



\subsection{Monolayer analysis}
\subsubsection{Interfacial tension measurement}
Time-resolved interfacial tension of the flat oil–water interface was measured with a platinum/iridium Wilhelmy plate connected to an electrobalance (KSV Nima, Finland)~\cite{maikawa2021engineering}. The Wilhelmy plate was immersed in a solution comprising an oil phase (with 1mM DPhPC concentration ensuring a saturating interface~\cite{venkatesan2015adsorption}) and a lower aqueous phase with salts in a Petri dish, and the interfacial tension was recorded for 10 min. The experiment was repeated at least three times.



\subsubsection{Complex viscosity measurement}
Interfacial shear rheology of the oil–water interface was measured using a Discovery HR-3 rheometer (TA Instruments) with an interfacial double wall ring geometry comprising a Du Noüy ring made of platinum/iridium wires (CSC Scientific, Fairfax, VA, catalog no. 70542000)~\cite{maikawa2021engineering}. Before each experiment, the Du Noüy ring was rinsed with ethanol and water and flame treated to remove organic contaminants. The solution chamber consisted of a double-wall Couette flow cell with an internal Teflon cylinder and an external glass beaker. A time sweep was performed at a strain of 1\% (within the linear regime) and a frequency of 0.05 Hz (sufficiently low such that the effects due to instrument inertia will not be significant). Interfacial complex surface viscosity $\mu_s$ was measured 5 minutes following the creation of the oil-water interface, and then the surface viscosity $\mu_s$ was recorded for another 5 min. The solution includes 10mM DPhPC in oil with given salts in water phase. The experiment was was repeated at least three times. 


\begin{figure*}[!h]
\centering
\includegraphics[width=\linewidth]{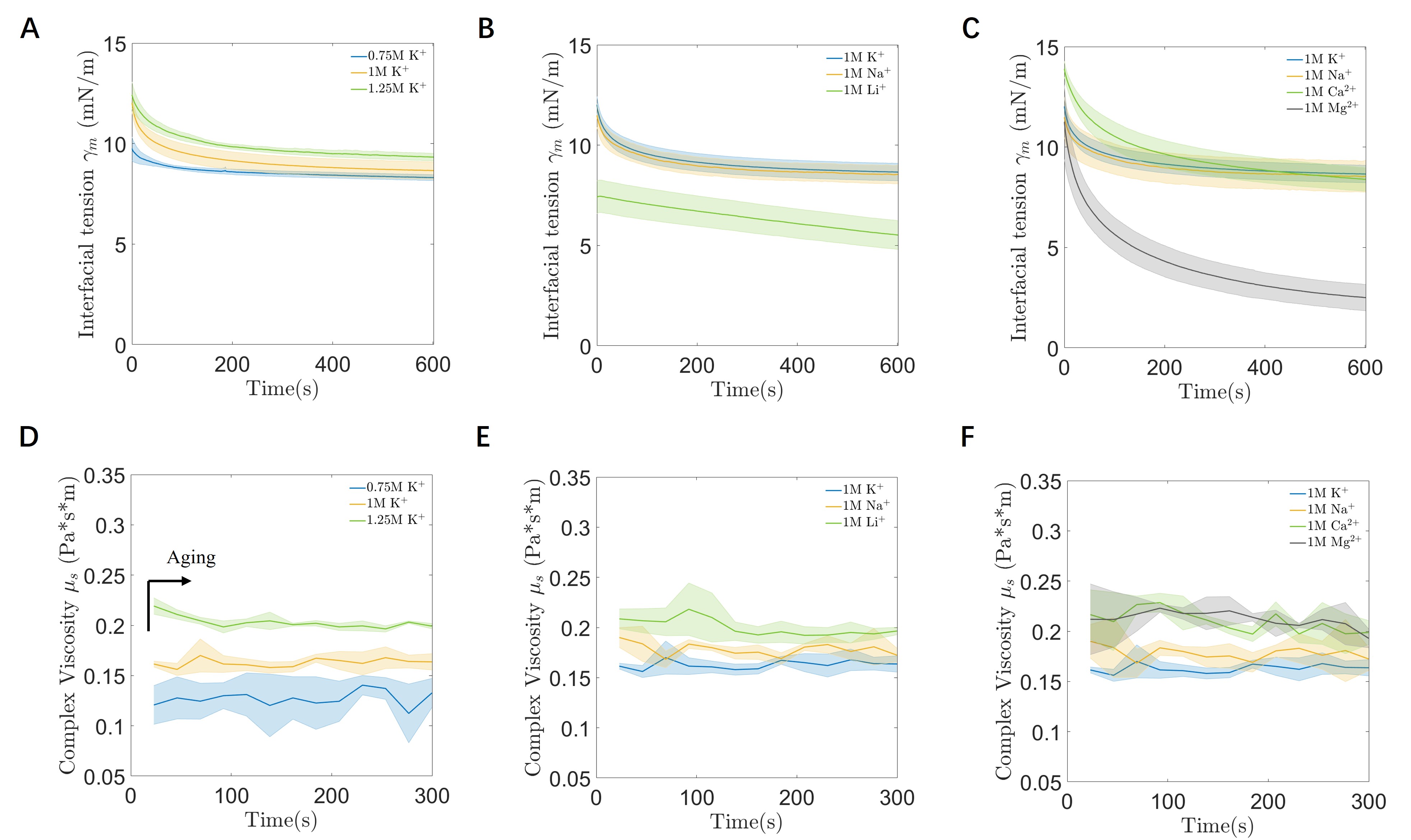} \caption{Salt effects on the interfacial tension (IFT, $\gamma_m$) (A-C) and complex viscosity $\mu_s$(D-F) for DPhPC monolayers.(A)IFT varied with KCl concentrations. (B)IFT varied with alkali ion size. (C) IFT varied with ion valence. (D)complex viscosity computed as a function of KCl concentrations. (E)complex viscosity computed as a function of alkali ion size. (F)complex viscosity computed as a function of ion valence. Complex viscosity results are recorded after 5 minutes of the aging time. Experimental error bars are included.} \label{fig:Monolayer}
\end{figure*}

\subsection{Bilayer analysis}
\subsubsection{Experimental setup to create DIBs}

A custom built setup named Interfacial Drainage Dilatational and Stability Stage (I-DDiaSS) is used to create and separate the droplet interface bilayers \cite{poulos2010automatable,bochner2017droplet,suja2020single}. As can be seen in Fig. \ref{fig:Setup}A, I-DDiaSS consists of a glass chamber, a blunt capillary needle (ID: 0.58 mm OD: 0.81 mm), a side camera (IDS UI 3060CP) and a motorized stage run by a stepper motor with a rotary encoder (Newport TRA12PPD and SMC100PP). The glass chamber holds the hexadecane solution and an aqueous sessile drop that is pinned at the center of the chamber bottom using a circular trench (0.3mm depth, not visible in the figure), and the sessile drop remains immobilized on the glass substrate during the course of the experiment. A capillary needle with a small droplet of agarose gel as a core at its tip holds the pendant drop in place \cite{leptihn2013constructing}. The side camera was used to obtain the drop profiles and bilayer radius using principles of shadowgraphy. Finally, a motorized stage below the glass chamber enables one to accurately position the sessile drop relative to a pendant drop to create or to separate the bilayers.


\subsubsection{Experimental protocol of DIBs} \label{sec:expprotocol}
At the start of an experiment a sessile drop with 3 $\upmu$L of salt solution was placed onto the circular trench on the bottom of the chamber, and was fully covered by 0.2 mL of the hexadecane solution with the DPhPC lipid. Then, a pendant drop of volume 1 $\upmu$L (same salt solution as the sessile drop) was added onto the capillary with the agarose core, which was then placed above the sessile drop. To study the salt effect on the mechanics of DIBs, we firstly use a KCl solution with  concentrations of 0.75M, 1M and 1.25M. Experiments were also conducted using 1M of NaCl, LiCl, CaCl$_2$ and MgCl$_2$, respectively.


To generate the lipid bilayers, the pendant drop is translated downward by the motorized stage into the lipid oil phase toward the sessile drop. Then both sessile and pedant drops are aged for 10 min to allow lipid monolayers to form at the surface of the droplets \cite{leptihn2013constructing}. After aging the motorized stage is used again to slowly push the sessile drop against the pendant drop for approximately 0.25 mm and then held in place \cite{najem2015activation}. The two monolayers form a bilayer when the thin liquid film between the pendant and sessile droplets drains away. As the bilayer forms an advancing front line can be observed between the droplets after a short period of time (see Fig. \ref{fig:Setup}C). For all the reported experiments, the initial bilayer radius was 0.20 mm. 


To conduct the bilayer separation experiments, the motorized stage is pulled downward, causing a traction force between the sessile and pendant drops. This process is carried out in a stepwise manner at a velocity of 0.05 mm/s for one second. The step size ($d$) has a constant value of 0.05 mm, resulting in step strain of $d/R_a = 0.067$ to extend the droplets, where $R_a = 0.75$ mm is the curvature at the apex of the pendant drop. The bilayer is allowed to relax for 90 seconds after each separation step until the sessile and pendant drops separate completely. We capture the entire process of the bilayer formation and separation using a side camera and analyzed the dynamics utilizing MATLAB. All experiments in this paper were performed at room temperature. Contributions from the mechanical compliance are ignored due to the relatively small strains acting on the droplets over the course of the experiment. A schematic diagram of the above mentioned bilayer formation and separation on the I-DDiaSS, as well as images obtained during these processes are shown in Fig. \ref{fig:Setup}B, C and D.


\subsubsection{Mathematical model of DIB separation} \label{sec:SimMethod}
Here we use a simple model based on the Young-Laplace equation for capturing the salt effects on the separation of the droplet interface bilayers. In this simplified model the agarose attached to the needle is ignored (See Fig. \ref{fig:SimFlowChart}A), and the drop profiles are further assumed to be symmetric to the plane of the bilayer, which holds true at low Bond numbers for sessile and pendant drops of comparable sizes. Under these assumptions, the non-dimensional Young-Laplace equation that governs the shapes of the drops can be written as, 

\begin{align}
    \frac{d\phi}{d\bar{s}}&=2-\text{Bo}\bar{z}-\frac{\sin{\phi}}{\bar{r}}\\
    \frac{d\bar{r}}{d\bar{s}}&=\cos{\phi}\\
    \frac{d\bar{z}}{d\bar{s}}&=\sin{\phi}
\end{align}

\begin{figure*}[!h]
\centering
\includegraphics[width=\linewidth]{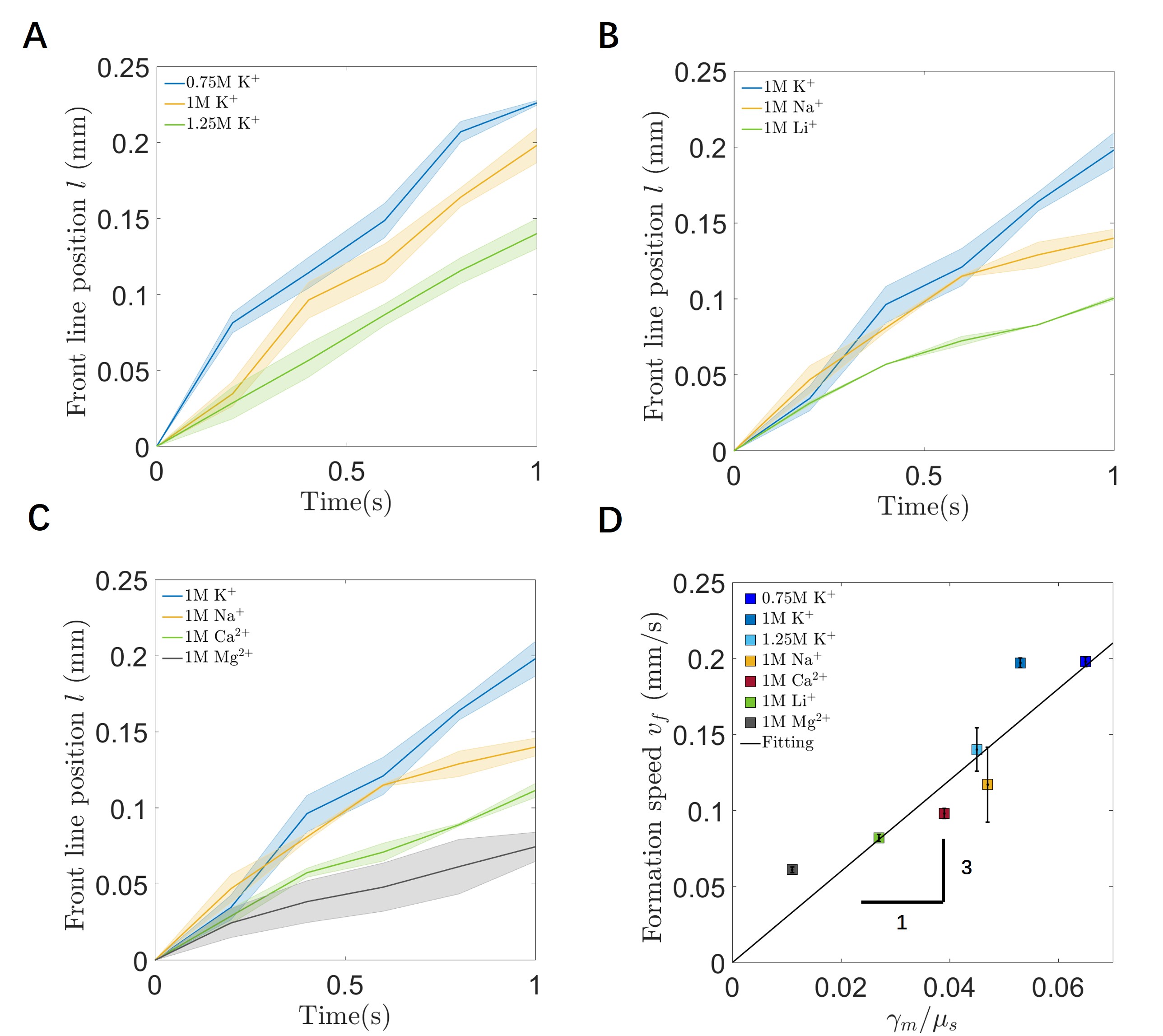} \caption{Bilayer front line position $l$ with the evolution of time (A)Front line position as a function of KCl concentrations. (B)Front line position as a function of alkali ion size. (C)Front line position as a function of ion valence. (D)Formation speed versus the ratio of IFT and complex viscosity. The solid line is a linear fit to the data and has a slope of $3$ with a goodness of fit $SSE=0.036$.} \label{fig:Formation}
\end{figure*}

\noindent where $\bar{s}$ is defined as the arc length along the pendant drop that is measured from the edge of the bilayer and non-dimensionalized by curvature of drop at the apex ($R_a$). $\phi$ is the angle between the tangent to the pendant drop profile and the horizontal. $\bar{r}$ and $\bar{z}$ represent the non-dimensional cylindrical coordinates of the bilayer interface. Bo is the Bond number denoted by Bo $=\Delta \rho g R_a^2/\gamma_m$, where $\gamma_m$ is the surface tension for the monolayer, and  $\Delta \rho$ is the density difference between aqueous and oil phases. For low Bond number drops having similar sizes, we can set the bilayer radius by the following force balance in the vertical direction \cite{huang2021surface}:  

\begin{equation} \label{eq:forcebalance}
   -\bar{R}_b^2 +\bar{F}_{ap} +  \bar{R}_b\sin{\theta_b} =0
\end{equation}
where $\bar{R}_b$ is the normalized bilayer radius, $\bar{F}_{ap}$ is the external force acting on the drop non-dimensionalized by $2\pi R_a \gamma_m$, $\gamma_m$ is the monolayer surface tension and $\theta_b$ is the contact angle between the monolayer and bilayer as illustrated in Fig. \ref{fig:SimFlowChart}A.  Physically, the first term is the non-dimensional Laplace pressure multiplied by the bilayer area originating from the deformation of the drops along the plane of the bilayer, the second term is the external force pushing the drops against each other and the third term is the non-dimensional interfacial tension multiplied by the circumference of the bilayer.

Finally the excess vertical interfacial tension acting on the bilayer, $\gamma_\perp - \gamma_{\perp,eq}$, can be related to the bilayer radius $R_b$ and the separation velocity of the bilayer, $v = -\frac{dR_b}{dt}$, where $\gamma_\perp =\gamma_m \sin \theta_b$ and $\gamma_{\perp,eq}$ is the value of $\gamma_{\perp}$ when $v = 0$. Expressing $\gamma_\perp$ in terms of the variables in Fig. \ref{fig:SimFlowChart}A closes the system of equations and gives the following non-dimensional evolution equation for $v$,

\begin{equation}\label{eq:SeperationVelocity}
    \bar{v}  = \frac{\bar{b}(\sin \theta_b - \sin \theta_{b,eq})}{K \bar{R_b}},
\end{equation}

where, $\bar{v} = \frac{\mu v}{\gamma_m}$ is the non-dimensional separation velocity, $\mu$ the bulk viscosity of the ambient oil phase,  $\theta_{b,eq}$ is the value of $\theta_b$ when the bilayer is in equilibrium during the separation, $\bar{b}$ the non-dimensional thickness of the bilayer and $K$ is a prefactor in the relationship between the excess force acting on the bilayer and the peeling rate of the bilayer.


We solve the system of equations above in an iterative process as illustrated in Fig. \ref{fig:SimFlowChart}B. Initially, a droplet profile can be analyzed using the Young-Laplace equation given the relevant values of Bo, $R_b$ and the contact angle $\theta_b$. From this droplet profile and under certain $\gamma_m$ and $\mu$ determined by the salt solutions, $\theta_b$ and $R_b$ can be extracted to give the corresponding $\gamma_\perp$ and $v$. Then the bilayer radius in the upcoming iteration step can be obtained from $v$ to generate a new droplet profile. In order to match the protocol in our experiments the iterations proceed in steps of $0.05$ s for a total duration of $90$ s. After the contact angle of an evolving bilayer approaches an equilibrium value $\theta_{b,eq}$, we alter the contact angle to mimic the contact angle change following a pull-up (extension) of the droplets in the actual experiment. $\bar{F}_{ap}$ is recalculated and the algorithm is repeated after the extension. This iterative process proceeds until the bilayer radius tends to zero to reflect a complete separation of the bilayer. Further details regarding the model and its solution are available in reference \cite{huang2021surface}.


\section{Results and discussion}\label{sec:results}
We will show the physicochemical properties of lipid monolayers and bilayers under three types of salt conditions: 1) the effect of KCl concentration (0.75M, 1M and 1.25M), 2) the effect of ion sizes (1M of KCl, NaCl and LiCl), and 3) the influence of ion valence (1M of KCl, NaCl, CaCl$_2$ and MgCl$_2$). In this section we first present the influence of salt on the interfacial tension (IFT) and the complex surface viscosity of lipid monolayers. Subsequently we report the effects of salt on bilayer formation and separation using DIBs. 

\subsection{Interfacial tension and complex viscosity} \label{sec:IFT}
Interfacial tension of the monolayers, $\gamma_m$, and the complex surface viscosity $\mu_s$ are two important physical properties that are influence by salts, and which in turn affects the formation and separation of DIBs. The IFT and the complex viscosity of the DPhPC monolayer at flat oil-water interfaces for different salts are reported in Fig. \ref{fig:Monolayer} A-C and Fig. \ref{fig:Monolayer} D-F, respectively. The shaded error bars indicate the standard errors obtained from three independent measurements.

At first sight, we see that the IFT for KCl, NaCl and CaCl$_2$ decays and then approaches an equilibrium value slightly below 10 mN/m, while for LiCl and MgCl$_2$ the value goes down slowly and reaches a value around and below 5 mN/m at 600 s. Taking the interfacial tension at 600 s as the equilibrium tension for every salt solution, we find that $\gamma_m$ at equilibrium increases with increasing KCl concentration. There is also a dependence on the size and valence of the ions where the IFT increases according to the following ranking K$^+$>Na$^+$>Ca$^{2+}$>Li$^+$>Mg$^{2+}$, indicating an increase with the alkali ion size and a decrease with the ion valence. Regarding the complex viscosity, we see that for all salts the value of $\mu_s$ stays constant versus time, which indicates that the systems have been in equilibrium before the measurement (data recorded 5 minutes following the formation of the oil-water interface). We also see that $\mu_s$ for all salts ranges from 0.1 to 0.25 Pa*s*m, which yields a Boussinesq number to be of the order of 10$^{5}$, suggesting a very viscous interface. Magnitudes of inferfacial viscosities of the order of 0.1 Pa*s*m are archievable due to the very low interfacial tensions that reflect densely packed monolayers as described by Raghunandan et al \cite{raghunandan2018predicting}. As seen in Fig. \ref{fig:Monolayer}D, the surface viscosity increases with increasing KCl concentration. Furthermore, inspection of Fig. \ref{fig:Monolayer}E indicates the surface viscosity increases with the alkali ion size. Finally, Fig. \ref{fig:Monolayer}F suggests that divalent ions produce larger surface viscosities than monovalent ions.


The above observations indicate that the ion concentration, ion size and ion valence influence the IFT and surface viscosities of the lipid monolayer. When ions in electrolyte solution approach and bind to the lipid, the lipid packing, lipid ordering, orientation of the lipid head group and the charge distribution are modified \cite{redondo2014structural,lin2016effects}. These measurements of interfacial properties on the DPhPC monolayer support the analysis of the formation and the separation of the DIB (See Section \ref{sec:formation}, \ref{sec:separation} and \ref{sec:simulation}). More investigations can be done to systematically and thoroughly study the mechanisms of the salt to the IFT and viscosity for lipid monolayers by experiment and simulations.

\subsection{Bilayer formation}\label{sec:formation}

\begin{figure*}[!h]
\centering
\includegraphics[width=\linewidth]{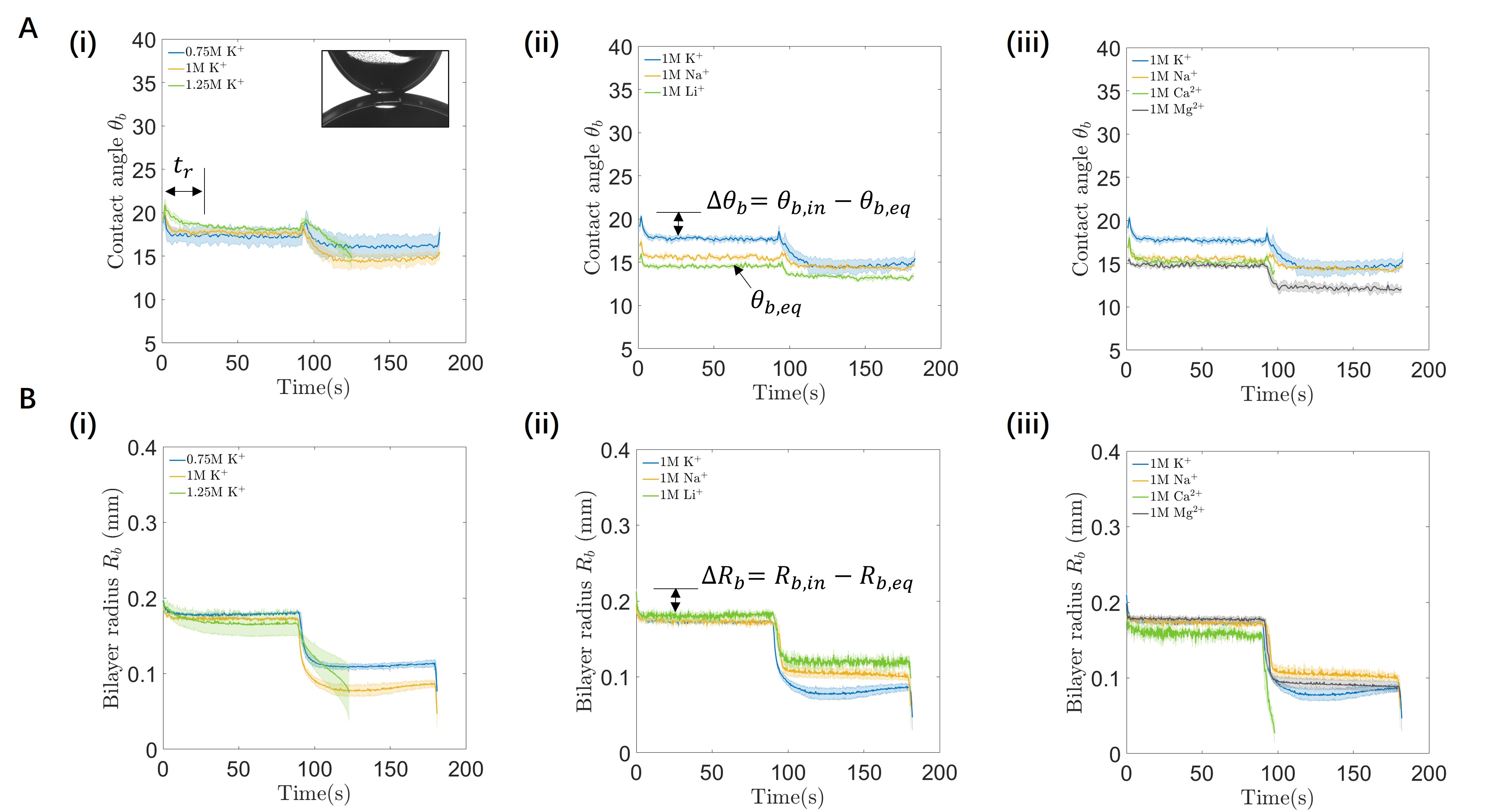} \caption{Mechanics of bilayer separation by salt effects. (A)(i)Evolution of the contact angle $\theta_b$ for 0.75M, 1M and 1.25M KCl solutions. (ii)Evolution of the contact angle $\theta_b$ for 1M of KCl, NaCl and LiCl. (iii)Evolution of the contact angle $\theta_b$ for 1M of KCl, NaCl and CaCl$_2$ and MgCl$_2$. (B)(i)Evolution of the bilayer radius $R_b$ for 0.75M, 1M and 1.25M KCl solutions. (ii)Evolution of the bilayer radius for 1M of KCl, NaCl and LiCl. (iii)Evolution of the bilayer radius for 1M of KCl, NaCl and CaCl$_2$ and MgCl$_2$. $t_r$ in plot is the relaxation time, $\theta_{b,in}$ and $\theta_{b,eq}$ are initial and equilibrium contact angle, and $R_{b,in}$ and $R_{b,eq}$ are initial and equilibrium bilayer radius.} \label{fig:Separation}
\end{figure*}

Bilayer formation refers to the process of two monolayers interdigitating to form a bilayer at the droplet-droplet interface. In order to study how the salts can influence the formation speed, the position of the advancing front of the bilayer is plotted as a function of time. These are shown and plotted in Fig. \ref{fig:Formation}A for three concentrations of K$^+$ ions. Likewise the influence of alkali ion size is shown in Fig. \ref{fig:Formation}B and the influence of ion valence is shown in Fig. \ref{fig:Formation}C. These plots indicate the progression of the bilayer fronts is approximately linear in time, and the respective velocities, $v_f$, in each case is estimated from the slopes. This method of determining formation velocities is similar to that used by Thutupalli et al and Vargas et al \cite{thutupalli2011bilayer,vargas2014fast}. It is evident from Fig. \ref{fig:Formation}A-C that the front velocity decrease with increasing KCl concentration, increase with alkali ion size and decrease with ion valence. 

We find that the dynamics of the bilayer formation is governed by the balance of surface viscous stress to surface tension. Taking the surface viscous stress to be the surface viscosity times an surface velocity gradient, this balance leads to:
\begin{equation} \label{eq:formationscaling}
    \mu_s \frac{v_f}{L}\approx \gamma_m
\end{equation}
where $L$ is the length scale over which the velocity changes. This motivates the plot shown in Fig. \ref{fig:Formation}D where the formation velocity is plotted against ratio of $\gamma_m/\mu_s$. The linearly of this plot supports the simple scaling analysis offered by Eqn \ref{eq:formationscaling}.



\subsection{Bilayer separation}\label{sec:separation}
The separation mechanics of DIBs have been previously investigated by our group by subjecting the bilayers to successive upward step strains as the sessile drop is shifted downward \cite{huang2021surface}. In this section, we tracked the mechanical response of the bilayer to the applied step strains by measuring the contact angle $\theta_b$ and the bilayer radius $R_b$ (See Fig. \ref{fig:Setup}D).

\begin{figure}[!h]
\centering
\includegraphics[width=\linewidth]{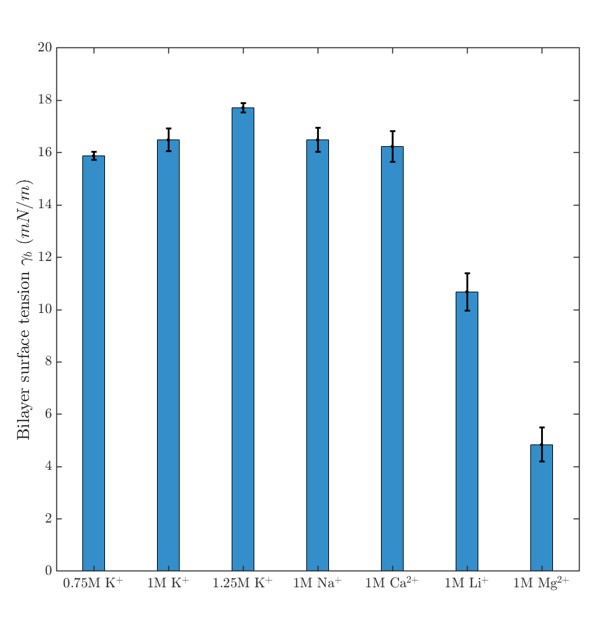} \caption{Bilayer surface tension $\gamma_b$ for seven salt conditions.} \label{fig:bilayertension}
\end{figure}

The evolution of the contact angle and bilayer radius of different salt environments are shown in Fig. \ref{fig:Separation}A and Fig. \ref{fig:Separation}B, respectively. Similar to our previous paper, these measurements at first sight reveal that for each salt sample the separation mechanics are primarily governed by the peeling processes, where in the peeling process we see the contact angle and bilayer radius decay from their initial values ($\theta_{b,in}$ or $R_{b,in}$) to the equilibrium values ($\theta_{b,eq}$ or $R_{b,eq}$). Examining the evolution of the contact angle and bilayer radius in response to successive extensions of strains of $d=0.067$, we observe that ultimately the radius rapidly drops to zero, which is an indication of a complete bilayer separation. Depending on the salt environment, separation occurs at either the first or the second extension step. One caveat is that based on the rule of a simple force balance (See Eqn \ref{eq:bilayerForce}), the equilibrium contact angle $\theta_{b,eq}$ of samples that have two successive extensions should be the same for the first and the second extensions. These measurement errors are due to optical artefacts relating to the droplet curvature at the interface. Looking into the equilibrium contact angle of the first extension in Fig. \ref{fig:Separation}A, we see that $\theta_{b,eq}$ increases with increasing KCl concentration, increases with alkali ion sizes but decreases with ion valence. Furthermore, we observe that in Fig. \ref{fig:Separation}A(i) the contact angle difference between the initial value after an extension and the equilibrium value, $\Delta \theta_b =\theta_{b,in}-\theta_{b,eq}$, increases with the KCl concentration. Inspections of Fig. \ref{fig:Separation}A(ii) indicates $\Delta \theta_b$ increases with the alkali ion size. Finally, Fig. \ref{fig:Separation}A(iii) suggests that there is a dependence on the size and valence of the ions at first extension, where the value increases following the order Ca$^{2+}$>K$^+$>Na$^+$>Mg$^{2+}$.

\begin{figure*}[!h]
\centering
\includegraphics[width=\linewidth]{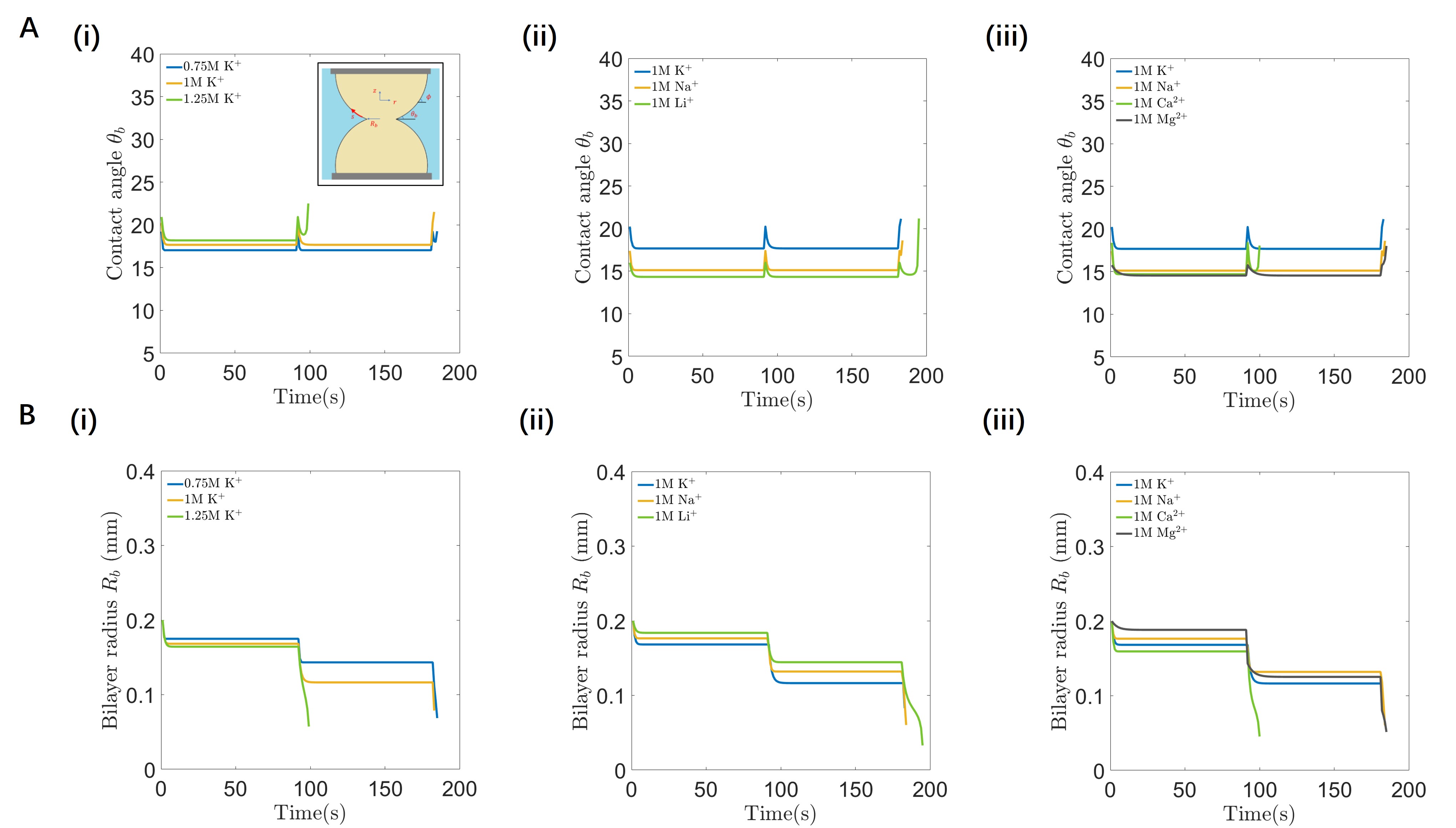} \caption{Results of salt effect on DIBs from simulation. (A)(i)Evolution of the contact angle $\theta_b$ by simulation for 0.75M, 1M and 1.25M KCl solutions. (ii)Evolution of the contact angle $\theta_b$ by simulation for 1M of KCl, NaCl and LiCl. (iii)Evolution of the contact angle $\theta_b$ by simulation for 1M of KCl, NaCl and CaCl$_2$ and MgCl$_2$. (B)(i)Evolution of the bilayer radius $R_b$ by simulation for 0.75M, 1M and 1.25M KCl solutions. (ii)Evolution of the bilayer radius by simulation for 1M of KCl, NaCl and LiCl. (iii)Evolution of the bilayer radius by simulation for 1M of KCl, NaCl and CaCl$_2$ and MgCl$_2$.} \label{fig:Simulation}
\end{figure*}

Regarding the salt effect on the evolution of the bilayer radius, we observe that the decay magnitude of the bilayer radius $\Delta R_b=R_{b,in}-R_{b,eq}$ after the first extension increases with the increase of the K$^+$ concentration (See Fig. \ref{fig:Separation}B(i)). In Fig. \ref{fig:Separation}B(ii) we see that $\Delta R_b$ following every extension increases with the alkali ion size. In Fig. \ref{fig:Separation}F we see that $\Delta R_b$ depends on ion size and valence, where the value increases following series Ca$^{2+}$>K$^+$>Na$^+$>Mg$^{2+}$>Li$^+$ at the first extension. Finally we estimate the relaxation time $t_r$ of the first step by fitting the decay of the contact angle with an exponential curve. Inspection of Fig. \ref{fig:Separation}A(i) and \ref{fig:Separation} B(i) indicates that $t_r$ at the first extension increases with increasing K$^+$ concentration and the rest of four sub-figures in Fig. \ref{fig:Separation} suggests that $t_r$ at the first extension increases following the order Mg$^{2+}$>Li$^{+}$>K$^+$>Ca$^{2+}$>Na$^+$. It's worth noting that except the 1.25M K$^+$ sample, the contact angle change in the last extension during the pulling mode is not obvious compared to the change for the bilayer radius, where a similar behavior is also found in our previous work \cite{huang2021surface}. For the 1.25M K$^+$ sample, we find that the change of the $\theta_b$ and $R_b$ in the last step is slower than the other salt samples, indicating that the last step of the sample consists of both the peeling and pulling process, where the peeling process dominates in the first 20 s.


The above observations indicate that salts can influence the evolution of the contact angle and bilayer radius during the peeling processes. First, by respectively examining the $\theta_{b,eq}$ of the initial step in each plot of Fig. \ref{fig:Separation}A, we find that the value increases with the increase of the IFT measured in Section \ref{sec:IFT}. Utilizing $\theta_{b,eq}$ and $\gamma_m$, we can obtain the bilayer surface tension $\gamma_b$, an intrinsic property dictating the force interactions of bilayers given by \cite{huang2021surface}:
\begin{equation}\label{eq:bilayerForce}
    \gamma_b = 2 \gamma_m \cos \theta_{b,eq}
\end{equation}
By calculating the bilayer tension for all salts (Fig. \ref{fig:bilayertension}), we observe that $\gamma_b$ increases with increasing K$^+$ concentration and decreases with ion size and the ion valence. Second, the salt effect on $\theta_{b,in}$ can be related to the mechanical compliance of the droplet that is governed by IFT and viscoelastic properties when the droplets are extended by step strains. Under this hypothesis, we plot the contact angle difference of the first step over $\gamma_m$ and find that $\Delta \theta_b$ increases with increasing IFT for all ions except Ca$^{2+}$ (See Fig. S2 in supplementary material), indicating the effect given by IFT dominates for most of the salts.

The dependence of $\Delta R_b$ and $t_r$ to the interfacial properties, especially when comparing samples among ion size and ion valence, still remains unclear. At first sight, these two variables can be related to the governing equation of the peeling process by viscous dissipation shown in our previous work given by \cite{huang2021surface,frostad2014direct,chatkaew2009dynamics}:
\begin{equation}\label{eq:JohnModel}
    R_b\frac{d R_b}{dt} =-\frac{b\gamma_m}{\mu}(\sin{\theta_b}-\sin{\theta_{b,eq}})
\end{equation}
From this equation we see that $R_b$ and $t$ depend on the interfacial tension, combining the viscosity component and the contact angle. Combine Eqn \ref{eq:JohnModel} with Eqn \ref{eq:forcebalance} we can further determine $\Delta R_b$ and $t_r$, which will be introduced and discussed in Section \ref{sec:simulation}. Finally when $R_b \rightarrow 0$, the viscous resistance becomes negligible close to bilayer separation, and contributions from lubrication forces dominate as peeling gives way to the pulling mode of separation \cite{frostad2014direct}.


\begin{figure*}[!h]
\centering
\includegraphics[width=\linewidth]{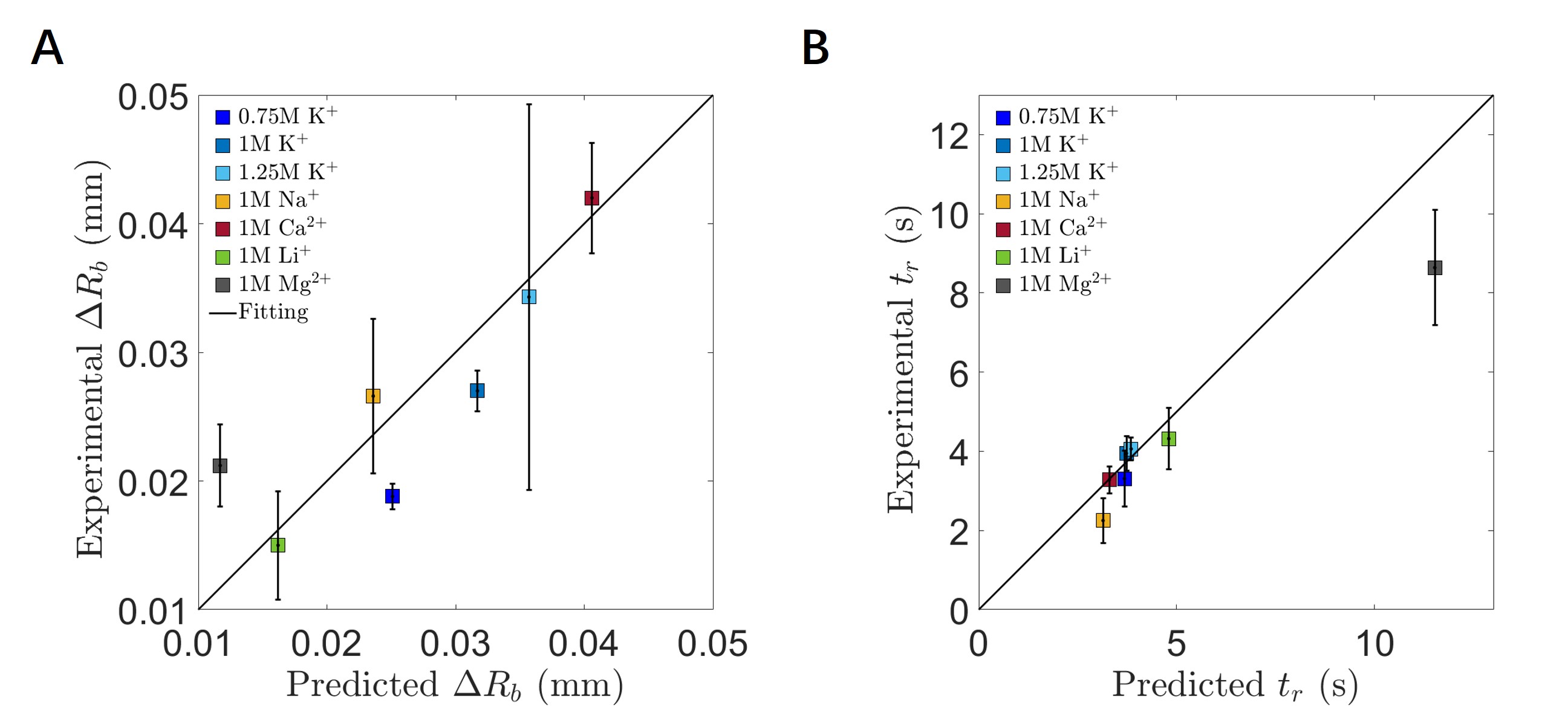} \caption{Comparison of experimental and simulated results of $\Delta R_b$ and $t_r$ at the first extension of the drops during the bilayer separation.(A) Predicted $\Delta R_b$ versus experimental value. (B)Predicted $t_r$ versus experimental value. Data close to solid lines($y=x$) indicates the predicted data is similar to the experimental data.} \label{fig:Fitting}
\end{figure*}

\subsection{Bilayer simulation}\label{sec:simulation}
In this section we report the predictions using the mathematical model described in Section \ref{sec:SimMethod}. In the simulation, values for $\theta_{b,in}$, $\theta_{b,eq}$ and $\gamma_m$ are found from measurements. The parameters Bo and $K$ are fitting parameters used to compare the simulations with the experiments. 

By solving the bilayer separation model we obtain the evolution of the contact angle (Fig. \ref{fig:Simulation}A) and the bilayer radius (Fig. \ref{fig:Simulation}B) how they are affected by KCl concentration, alkali ion size and ion valence. First, we see that the model successfully simulates the relaxation of $\theta_b$ and the decay of $R_b$ for each sample during the separation. Second, in Figs. \ref{fig:Simulation}B(i) and \ref{fig:Simulation} B(ii) we observe that $\Delta R_b$ increases with increasing KCl concentration and the alkali ion sizes. Third, Fig. \ref{fig:Simulation}B (iii) suggests that the decay of the bilayer radius increases following the order Ca$^{2+}$>K$^+$>Na$^+$>Mg$^{2+}$. Regarding the relaxation time we find that $t_r$ increases with the KCl concentration, and increases following the order Mg$^{2+}$>Li$^{+}$>K$^+$>Ca$^{2+}$>Na$^+$. We also extract the predicted bilayer radius and relaxation time of the first extension, and then plotted them with the experimental ones (See Fig. \ref{fig:Fitting}). As can be seen, the predicted $R_b$ is close to the experimental value, and the predicted $t_r$ of most salts are similar to the experimental data, and the $t_r$ of Mg$^{2+}$ by experiment is slightly lower than the simulated result. 




The change of the decay magnitude of the bilayer radius, as well as the relaxation time across subsequent separation steps for different salt conditions are intrinsic features of DIB separation under step strain. Utilizing the framework of our previous work we can obtain the expression of a normalized bilayer radius difference, $\Delta \bar{R_b}=\Delta R_b/R_a$, as \cite{huang2021surface}:

\begin{align}\nonumber \label{eq:radiussine}
\Delta \bar{R}_b &= \frac{1}{2} (\sin \theta_{b,in} -\sin \theta_{b,eq})\\ 
&+ \frac{1}{8\bar{R}_{b,in}} (\sin^2 \theta_{b,in} -\sin^2 \theta_{b,eq}) + \mathcal{O}\left(\frac{\sin^3 \theta_b}{\bar{R}_{b,in}^2}\right)
 \end{align} 
We see that $\Delta \bar{R}_b$ scales with ($\sin \theta_{b,in} -\sin \theta_{b,eq}$) and ($\sin^2 \theta_{b,in} -\sin^2 \theta_{b,eq}$). Neglecting the third term of $\Delta \bar{R}_b$ and dividing it by the average velocity obtained from Eqn \ref{eq:SeperationVelocity} we can estimate the relaxation time given by:
\begin{equation} \label{eq:trsine}
    \bar{t}_r=\frac{ K}{16 \bar{b}\bar{R}_{b,in}}(\sin \theta_{b,eq} +\sin \theta_{b,in}+4\bar{R}_{b,in})
\end{equation}
Clearly, we can see that the relaxation time depends on the prefactor $K$, the contact angle and the bilayer radius.

\section{Conclusion}

In this manuscript we studied the salt effects to the monolayers at the flat oil-water interfaces and on the formation and separation of droplet interface bilayers. We showed that:
1)the IFT of the monolayer at flat oil-water interface increases with KCl concentration and alkali ion sizes but decreases with ion valency.
2)the surface viscosity of the the monolayer at flat oil-water interface increases with KCl concentration and ion valency, but decreases with alkali ion size.
3)the formation speed of the droplet interface bilayer scales with the ratio of the IFT to the surface viscosity.
4)the equilibrium contact angle during the DIB separation and the bilayer surface tension increase with KCl concentration and alkali ion sizes but decrease with ion valence.
5)the contact angle difference during the DIB separation increases with KCl concentration and alkali ion size, and depends on ion valency.
6)the bilayer radius and the relaxation time during the separation of DIB depend on the salt concentration, ion size and ion valence.
We also reported a simple mathematical model that successfully simulates the separation process and captures the features mentioned above. These results improve our understanding of how salts affect the interfacial and rheological properties of a lipid monolayer, as well as its influence the formation and separation mechanics of the bilayers under step strain.


There remain several opportunities for future work. First, future studies may include the effect of anions such as Cl$^-$, Br$^-$ and I$^-$, as these halide anions can interact with positively charged groups for some phospholipids thus changing there physicochemical properties \cite{lin2016effects}. Second, it would be worthwhile to study the separation mechanics under constant separation rates or under a constant separation force \cite{frostad2014direct,brochard2003unbinding}. Finally, investigating the salt effect on the pulling mode during the bilayer separation is also a promising direction for future research.

\section{Supplementary material}
See the supplementary material for details on the data acquisition of the DIB profile, the formation speed versus the ratio of the IFT to bulk viscosity, the salt effect to the contact angle difference and the derivation of Eqn \ref{eq:radiussine} and \ref{eq:trsine}.

\section{Authors' Contributions}
Y. H. conceived the study, designed and performed experiments, developed the algorithms, analyzed the data and wrote the manuscript. V.C.S conceived the study, analyzed the data and wrote the manuscript. L.A. performed the experiments. G.G.F. conceived and supervised the study, designed the experiments and critically reviewed the
manuscript.

\section{Acknowledgements}
We thank Aadithya Kannan for assistance in running the experiments, and the reviewers for comments and suggestions that led to significant improvement of the paper.

\section{Declaration of Competing Interest}
The authors declare that they have no known competing financial interests that could have appeared to influence the work reported in this paper.

\section{Data availability statement}
The data that support the findings of this study are available from the corresponding author upon reasonable request.

\bibliographystyle{vancouver}
\bibliography{References}

\end{document}